\documentclass[aps,pre,twocolumn,superscriptaddress,amsmath,amsfonts,showpacs]{revtex4-1}
\usepackage{amsmath}
\usepackage{amssymb}
\usepackage{graphicx}
\usepackage{subfigure}
\usepackage{mathrsfs}
\usepackage{hyperref}
\usepackage{color}

\begin{document}

\title{Phase noise of oscillators with unsaturated amplifiers}%
\author{Eyal Kenig}
\email[Corresponding author:\ ]{eyalk@caltech.edu}
\affiliation{Department of Physics, California Institute of Technology, Pasadena, California 91125, USA}
\author{M. C. Cross}
\affiliation{Department of Physics, California Institute of Technology, Pasadena, California 91125, USA}
\author{Jeff Moehlis}
\affiliation{Department of Mechanical Engineering, University of California, Santa Barbara, California 93106, USA}
\author{Kurt Wiesenfeld}
\affiliation{School of Physics, Georgia Institute of Technology, Atlanta, GA 30332, USA}

\begin{abstract}
We study the role of amplifier saturation in eliminating feedback noise in self-sustained oscillators. We extend previous works that use a saturated amplifier to quench fluctuations in the feedback magnitude, while simultaneously tuning the oscillator to an operational point at which the resonator nonlinearity cancels fluctuations in the feedback phase. We consider a generalized model which features an amplitude-dependent amplifier gain function. This allows us to determine the total oscillator phase noise in realistic configurations due to noise in both quadratures of the feedback, and to show that it is not necessary to drive the resonator to large oscillation amplitudes in order to eliminate noise in the phase of the feedback.
\end{abstract}
\date{\today}


\maketitle

\section{Introduction}  Some time ago, Greywall $\textit{et al.}$ demonstrated an interesting noise quenching effect in the operation of a self-oscillating system \cite{Greywall_PRL, Greywall_PRA}, a discovery that has important potential impact for the design of high frequency, low noise electronic oscillators \cite{Vig,Villanueva}. In addition to its practical consequences, the noise quenching phenomenon is of fundamental interest because it appeared when the system operated in the nonlinear regime, {\it i.e.} the quenching apparently relies on the inherent nonlinearity of the resonator.  In fact, the authors drew a connection between the optimal operating point (from the perspective of noise quenching) and a bifurcation point of the associated ``open loop" system, the so-called cusp point of the driven damped Duffing oscillator.  This connection is counter-intuitive since, quite generally, one associates bifurcation points with enhanced noise sensitivity.  More recently, Kenig $\textit{et al.}$ \cite{Kenig_PRL, Kenig_PRE} showed that this phenomenon is not restricted to the specific Duffing-like system studied by Greywall {\it et al.}, by reformulating the dynamical problem in a more general setting.

The purpose of this paper is to explore the conditions under which perfect phase noise quenching can be achieved.  We reconsider the system of Ref.~\cite{Greywall_PRL} using a generalized, more realistic amplifier feedback term, having both a low amplitude linear gain regime and a large amplitude saturated gain regime.  We find, first, that complete noise quenching only occurs when the system is operated in the high amplitude, fully saturated regime, which corresponds to the system studied in Ref.~\cite{Greywall_PRL}.  We also show that substantial phase noise reduction can be achieved away from this limit, even when the oscillator operates at amplitudes far below the critical point for bifurcations of the associated open loop system.
Going from a saturated to an unsaturated amplifier leads to two different effects. Firstly, since the drive on the resonator is no longer constant as parameters such as the phase of the feedback signal are changed, the closed loop oscillator behavior is no longer simply related to the open loop resonator response curves, and the optimal operating points are not given by the turning points of the resonator Duffing curve. Secondly, noise from the amplifier is no longer confined purely to the phase direction, since fluctuations in the magnitude of the drive are no longer quenched by the saturation.

\section{Model Equations and Phase Space View}  Greywall $\textit{et al.}$ considered a nonlinear resonator typified by a thin, electrically conducting beam of mass $M$ in a uniform magnetic field and driven by an alternating feedback current \cite{Greywall_PRL, Greywall_PRA}.  The beam dynamics are governed by the equation \cite{Greywall_PRA}
\begin{equation}\label{}
    M\ddot{X} + \mu \dot{X} + K_1 X + K_3 X^3 = F,
\end{equation}
where $X$ is the beam's displacement from equilibrium, $\mu$ is the damping coefficient, $K_1$ and $K_3$ are the linear and nonlinear restoring force parameters, respectively, and $F$ is the feedback force provided by a series amplifier/phase-shifter/limiter combination.  Here, the amplifier boosts the signal to overcome dissipation, while the phase-shifter introduces a readily accessible control parameter to tune the system to the desired operating point.  The functional purpose of the limiter is not obvious, but as we'll see, it plays an important role in eliminating input noise generated by the amplifier.

\begin{figure}[]
\begin{center}
  \includegraphics[width=1\columnwidth]{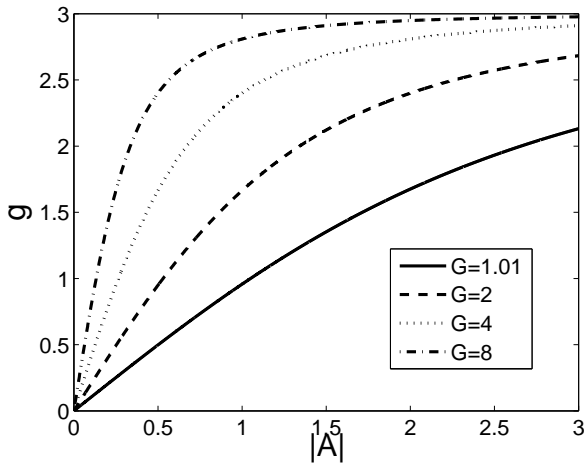}
  \caption{\label{fig1} The amplifier profile for $k=2$ and $s=3$.}
\end{center}
\end{figure}

In the high-$Q$ limit, the system dynamics are well described by a slowly varying complex amplitude $A$.  By appropriate scaling, the deterministic evolution of $A$ is described by
\begin{equation}\label{ampEq}
    \frac{dA}{dT} = \left( - \frac{1}{2} + i \frac{3}{8} \left|A\right|^2 \right) A + \frac{{\cal H}(A)}{2} e^{i \Delta},
\end{equation}
where $T$ is the slow time, ${\cal H}$ represents the action of the amplifier and $\Delta$ that of the phase shifter.  The amplifier is assumed to affect only the magnitude of $A$, with ${\cal H}$ having the form
\begin{equation}\label{}
    {\cal H}(A) = g(|A|) \, \frac{A}{|A|} \, .
\end{equation}
Refs.~\cite{Greywall_PRL, Greywall_PRA} studied the case of a strictly saturated amplifier, which corresponds to the case where $g$ is a constant.  Here, we consider the more general Rapp Model~\cite{Rapp,RappBook}, widely used in the engineering literature for solid state power amplifiers, which includes a non-saturated regime:
\begin{equation}\label{Rapp}
     g(|A|) = \frac{G|A|}{\left[1 + \left(\frac{G|A|}{s}\right)^k \right]^{1/k}},
\end{equation}
where $G$ and $s$ are constants, and $k$ is an integer which controls the crossover between the low amplitude, linear gain regime ($g \sim G|A|$) and the large amplitude, saturated regime ($g \sim s$).  In particular, we recover the strict saturation case by taking the limit $k \to \infty$ and/or $G \to \infty$.  Figure \ref{fig1} plots $g(|A|)$ for some typical parameters.

Although noise can enter the system in a variety of ways, we want to focus on the noise quenching phenomenon originally discussed in Refs.~\cite{Greywall_PRL, Greywall_PRA}, which specifically relates to the amplifier noise.
Amplifier noise will in general have components in both the magnitude and phase quadratures, which can be accounted for in the amplitude equation by adding to the right-hand side of Eq.~(\ref{ampEq}) the complex noise $\Xi e^{i\Phi}e^{i\Delta}/2$ with $\Xi=\Xi_{R}+i\Xi_{I}$ and $\Phi$ the phase of the  amplitude $A$. This yields the noisy amplitude equation
\begin{equation}
\frac{dA}{dT} = \left( - \frac{1}{2} + i \frac{3}{8} \left|A\right|^2 \right) A +\frac{1}{2}\left[g(|A|)\frac{A}{|A|} + \Xi e^{i\Phi}\right]e^{i\Delta} \, .
\label{noisy_amplitude_equation}
\end{equation}
For models of the amplifier noise we have investigated, and for this definition of $\Xi$ (with the phase factor involving $\Delta$ explicitly factored out), $\Xi_{R}$ and $\Xi_{I}$ are uncorrelated, and we will assume this is true in the remainder of this paper. The relative magnitude of $\Xi_{R}$ and $\Xi_{I}$ depend on the properties of the noise source and the saturation level of the amplifier.

\begin{figure}[]
\begin{center}
      \subfigure[]{
   \includegraphics[width=0.48\columnwidth]{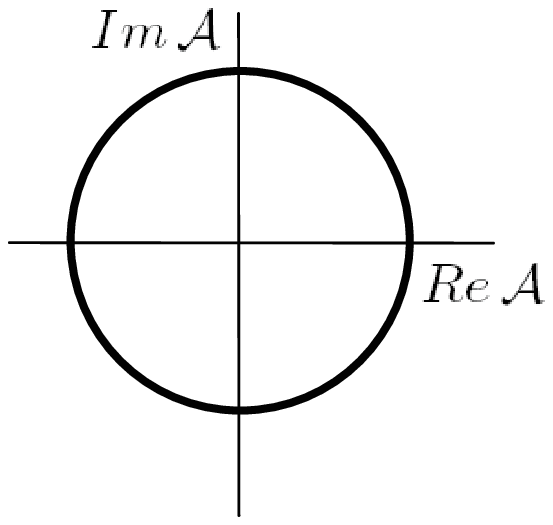}}
  \subfigure[]{
   \includegraphics[width=0.48\columnwidth]{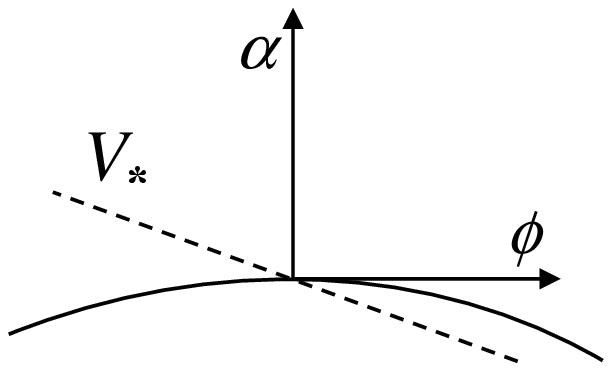}}
      \subfigure[]{
   \includegraphics[width=0.48\columnwidth]{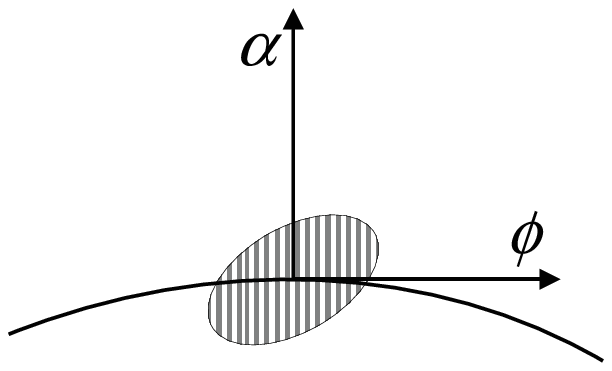}}
  \subfigure[]{
   \includegraphics[width=0.48\columnwidth]{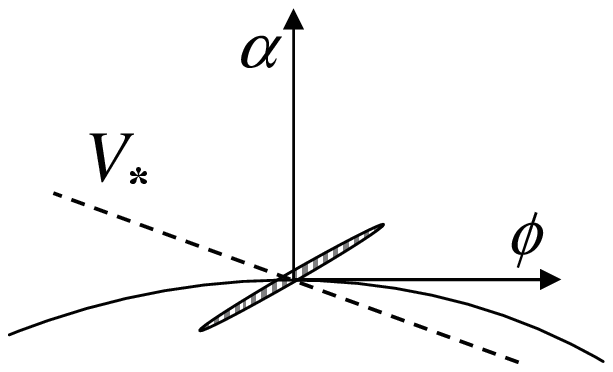}}
      \subfigure[]{
   \includegraphics[width=0.48\columnwidth]{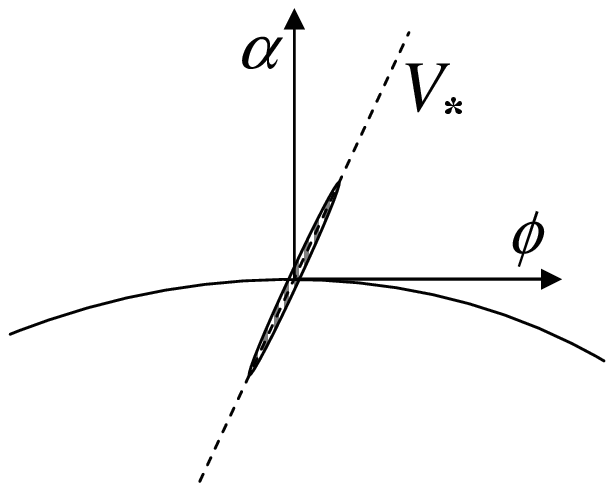}}
  \subfigure[]{
   \includegraphics[width=0.48\columnwidth]{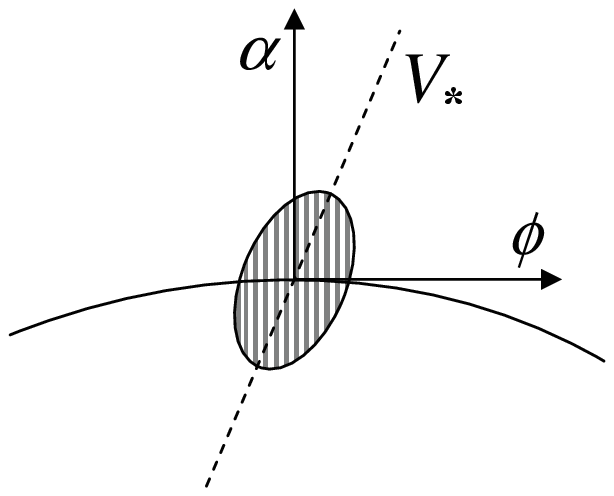}}
   \end{center}
   \caption{\label{fig2} Phase space of the complex amplitude $A$.  (a) In a rotating frame, the noise free orbit is circle of fixed points.  (b) Close up view showing local coordinate axes $(\alpha,\phi)$ and the vector $V_\star$ along which perturbations relax back to the original fixed point. (c) The hatched region represents the two dimensional (typically anisotropic) noise distribution.  (d) In the strict saturation limit, the noise cloud collapses to one dimension.  (e) Parameter tuning can align the noise with $V_\star$, thus eliminating phase diffusion.  (f) More generally, the noise cloud stays two dimensional, and parameter tuning results in partial reduction of phase diffusion.}
\end{figure}

Before turning to the calculations, it is worthwhile to consider the essential dynamics in a qualitative way, as captured by the phase space geometry of the system.  Figure~\ref{fig2} shows the situation in the complex-$A$ plane.  In the absence of noise, the system has an attracting orbit with uniform angular velocity, represented by the circle in Fig.~\ref{fig2} (a).  In a rotating frame, this becomes a circle of attracting fixed points.  Depending on the initial condition, the system trajectory ends up at one or another of the equilibria.  Suppose that the system has settled down to the particular point $x_0$, and consider the effect of an isolated perturbation:  the system is pushed off $x_0$, and subsequently relaxes back to some point on the circle.  Typically, the new fixed point is not $x_0$, and this corresponds to a net phase drift of the oscillator.  A single kick might advance the phase, or retard it; but there is a special set of perturbations (labeled $V_\star$ in Fig.~\ref{fig2} (b)) for which the system evolves back to $x_0$, resulting in no phase drift. These are perturbations
along the eigenvector of the linear flow with negative eigenvalue.

Instead of an isolated perturbation, noise has the effect of continually kicking the system trajectory.    Ordinarily, the individual noise kicks fluctuate in magnitude and direction, though perhaps not isotropically:  the oblong shaded region in Fig.~\ref{fig2} (c) represents the distribution of noise kicks.  Over time, the corresponding accumulation of phase shifts gives rise to phase diffusion.  But suppose the noise is confined to one dimension only (Fig.~\ref{fig2} (d)), and furthermore suppose that this direction coincides with $V_\star$ (Fig.~\ref{fig2} (e)).  Under these circumstances there would be no phase diffusion.  The quenched-noise phenomenon identified by Greywall {\it et al.} corresponds to this situation.  Physically, we'll show that it hinges on the strict saturation property of the amplifier, in addition to tuning the system parameters. From a calculational point of view, the vector perpendicular to $V_\star$ is of particular importance.  It plays a direct role in explicit determination of $V_\star$ on the one hand, and is central in generating complete expressions for the system's power spectrum.  We'll denote it by $V_\perp$.  Formally, $V_\perp$ is the null left eigenvector of the Jacobian; physically, it is the direction of maximum phase noise sensitivity.

\section{Fluctuation Evolution Equations}
Our starting point is the noisy amplitude equation (\ref{noisy_amplitude_equation}). For now we allow the spectral composition of $\Xi_R$ and $\Xi_I$ to be general, and assume only that the noise is weak enough that the system dynamics remain close to the deterministic limit cycle. Consider first the noise-free dynamics ($\Xi = 0$).  We substitute $A = a e^{i\Phi}$ into Eq.~(\ref{noisy_amplitude_equation}), divide out a common factor of $e^{i\Phi}$, and separate the real and imaginary parts to give
\begin{eqnarray}
\frac{da}{dT} = - \frac{a}{2} + \frac{g(a)}{2} \cos \Delta = f_a(a) \, , \\
\frac{d\Phi}{dT} = \frac{3}{8}a^2 + \frac{g(a)}{2} \frac{\sin \Delta}{a} = f_\Phi(a) \, . \nonumber
\label{noisefreearray}
\end{eqnarray}
We identify the amplitude $a_0$ of the periodic orbit by setting $f_a(a_0) = 0$, which yields
\begin{equation}
a_0 = \frac{s}{G} \left[ \left( G \cos \Delta \right)^k - 1 \right]^\frac{1}{k} \, .
\label{a0_solution}
\end{equation}
The corresponding frequency $\Omega_0$ is given by $\Omega_0 = f_\Phi(a_0)$.  The vector $V_\perp$ is \cite{Kenig_PRE}
\begin{equation}
V_\perp = \left( - \frac{\partial f_\Phi(a_0)/\partial a}{\partial f_a(a_0)/\partial a}, 1\right).
\label{Vperp}
\end{equation}

Now consider the effect of noise.  We set $a = a_0 + \alpha$ and $\Phi = \Omega_0 T + \phi$, with $\alpha, \phi$ small, and substitute into the governing Eq.~(\ref{noisy_amplitude_equation}).  We expand to first order in the small quantities, divide out a common exponential factor and separate real and imaginary parts as before, to get (omitting the algebra) the evolution equations for the perturbations
\begin{equation}
\frac{d\alpha}{dT} = -\frac{\alpha}{2} +  \frac{1}{2}\left\{ \cos \Delta \left(g^\prime(a_0)\alpha + \Xi_{R} \right) - \sin \Delta\,\Xi_{I} \right\},
\end{equation}
\begin{equation}
a_0\frac{d\phi}{dT} = -\alpha \Omega_0 + \frac{9}{8} a_0^2 \alpha +\frac{1}{2} \left\{ \cos \Delta \, \Xi_{I}  + \sin \Delta  \left( g^\prime(a_0)\alpha + \Xi_{R} \right) \right\} \, .
\end{equation}
Finally, recast this into vector form
\begin{eqnarray}
\frac{d}{dT} \left( \begin{array}{c} \alpha \\ \phi \end{array} \right) &=& \left( \begin{array}{lr} \frac{\partial f_a(a_0)}{\partial a}  & 0  \\ \frac{\partial f_\Phi(a_0)}{\partial a}  & 0 \end{array} \right) \left( \begin{array}{c} \alpha \\ \phi \end{array} \right) + \frac{\Xi_{R}}{2}  \left( \begin{array}{c} \cos \Delta \\ a_{0}^{-1} \sin \Delta \end{array} \right)\nonumber\\
&+& \frac{\Xi_{I}}{2}  \left( \begin{array}{c} -  \sin \Delta \\ a_{0}^{-1}\cos \Delta \end{array} \right) \, .
\label{vector_SDEs}
\end{eqnarray}
The relative strength of $\Xi_{R}$ and $\Xi_{I}$ can depend on aspects of the amplifier beyond those captured by its gain function for a periodic signal. However, as a simple model of the effect of amplifier saturation on this ratio we imagine, following Ref.~\cite{Greywall_PRA}, a noise source at the amplifier input, corresponding to the replacement (see Eq.(\ref{ampEq}))
\begin{equation}
{\cal H}(A) \to {\cal H}(A+\xi) \label{Simple_amplifier_noise} ,
\end{equation}
where the complex noise $\xi = \xi_R + i \xi_I$ with $\xi_R$ and $\xi_I$ real, equal intensity, and statistically independent. Physically, adding noise to the complex amplitude $A$ in this way corresponds to
considering noise passed through a narrow-band filter around the oscillation frequency before entering the amplifier. This captures the effect the amplifier has on noise in the frequency band near the carrier, though it ignores the up- and down-conversion of noise from the vicinity of other harmonics to the carrier frequency that would occur for wide band noise.
With the replacement (\ref{Simple_amplifier_noise}) the feedback function becomes
\begin{equation}\label{}
    g(|A+\xi|) \frac{A+\xi}{|A+\xi|}e^{i \Delta}.
\end{equation}
Expanding to linear order in $\xi$ and defining  $\bar\xi=e^{-i\Phi}\xi$, the term
\begin{equation}\label{}
    \frac{A+\xi}{|A+\xi|}\simeq e^{i\Phi}\left(1+i\frac{\bar\xi_{I}}{a_0}\right)\label{Simple Xi},
\end{equation}
represents noise in the phase of the feedback. The term $g(|A+\xi|)\simeq g(a_0)+g'(a_0)\bar\xi_{R}$ contains only fluctuations in the magnitude of the feedback.
This model therefore gives
\begin{equation}
\Xi_{R}=g'(a_{0})\bar\xi_{R},\quad\Xi_{I}=a_{0}^{-1}g(a_{0})\bar\xi_{I},
\label{XiRandXiI}
\end{equation}
with $\bar\xi=\bar\xi_{R}+i\bar\xi_{I}$. Note that $\bar\xi_{R},\bar\xi_{I}$ are again real, equal intensity, and statistically independent noise terms.
The expressions (\ref{XiRandXiI}) are consistent with the expected limits of equal strengths of $\Xi_{R}$ and $\Xi_{I}$ for a linear amplifier.

\section{Phase Noise Quenching}

\subsection*{Large Amplitude Limit}  We're now in a position to understand the origin of the noise quenching phenomenon. Equation (\ref{vector_SDEs}) governs the system's response to noise.  In the general case, because the noise functions $\Xi_{R}$ and $\Xi_{I}$ are independent, and these multiply vectors having different directions, the total input noise fluctuates over all directions in the phase space plane.  But in the large amplitude limit, the amplifier is saturated, and from Eq.~(\ref{XiRandXiI}) $\Xi_{R}\to 0$ since $g'\to 0$ in this limit, thus eliminating one noise term.  The remaining noise source fluctuates in magnitude but points along a fixed direction in phase space, and by tuning the system parameters one can arrange for this direction to be perpendicular to $V_\perp$:
\begin{equation}\label{}
    V_\perp \cdot \left( \begin{array}{cc} - \sin \Delta \\ a_0^{-1} \, \cos \Delta \end{array} \right) = 0  \, ,
\end{equation}
which guarantees the elimination of phase diffusion.  This condition takes the explicit form (upon evaluating Eq.~(\ref{Vperp}) in the limit $G \to \infty$)
\begin{equation}
1=\frac{3}{2} s^2 \cos^3\Delta \sin \Delta.
\label{saturated_quenching_condition}
\end{equation}
This condition has solutions for $\Delta$ providing $s>(4/3)^{5/4}$. Note that this condition is identical to the condition
for having bi-stability in the open loop system driven at the saturation value $s$ \cite{LCreview}.
In general, except in this limit of saturated amplifier output, the input noise is not confined to a fixed direction, and no amount of tuning can fully eliminate the phase diffusion.

\subsection*{Unsaturated Regime}
\begin{figure}[]
\begin{center}
      \subfigure[]{
   \includegraphics[width=1\columnwidth]{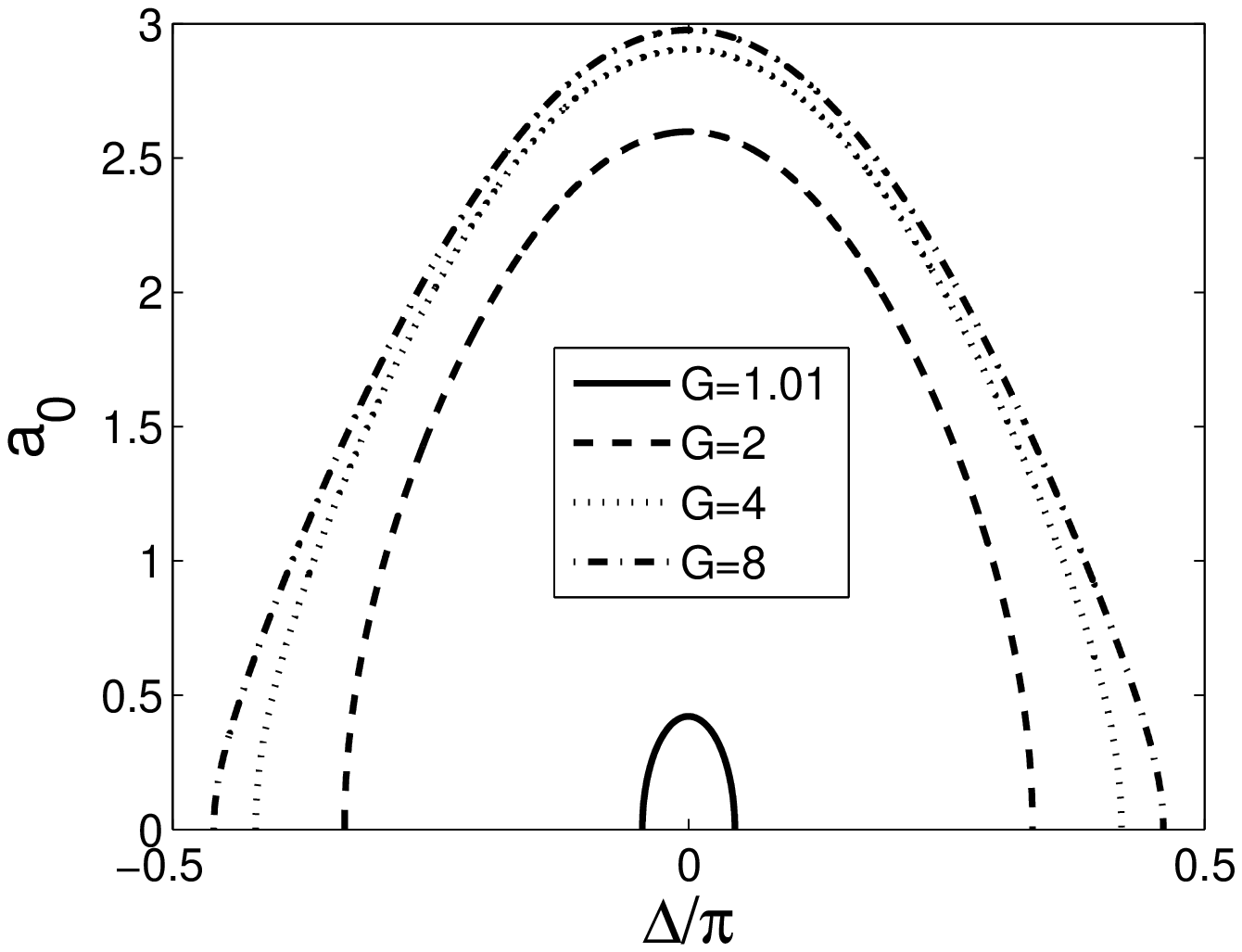}}
  \subfigure[]{
   \includegraphics[width=1\columnwidth]{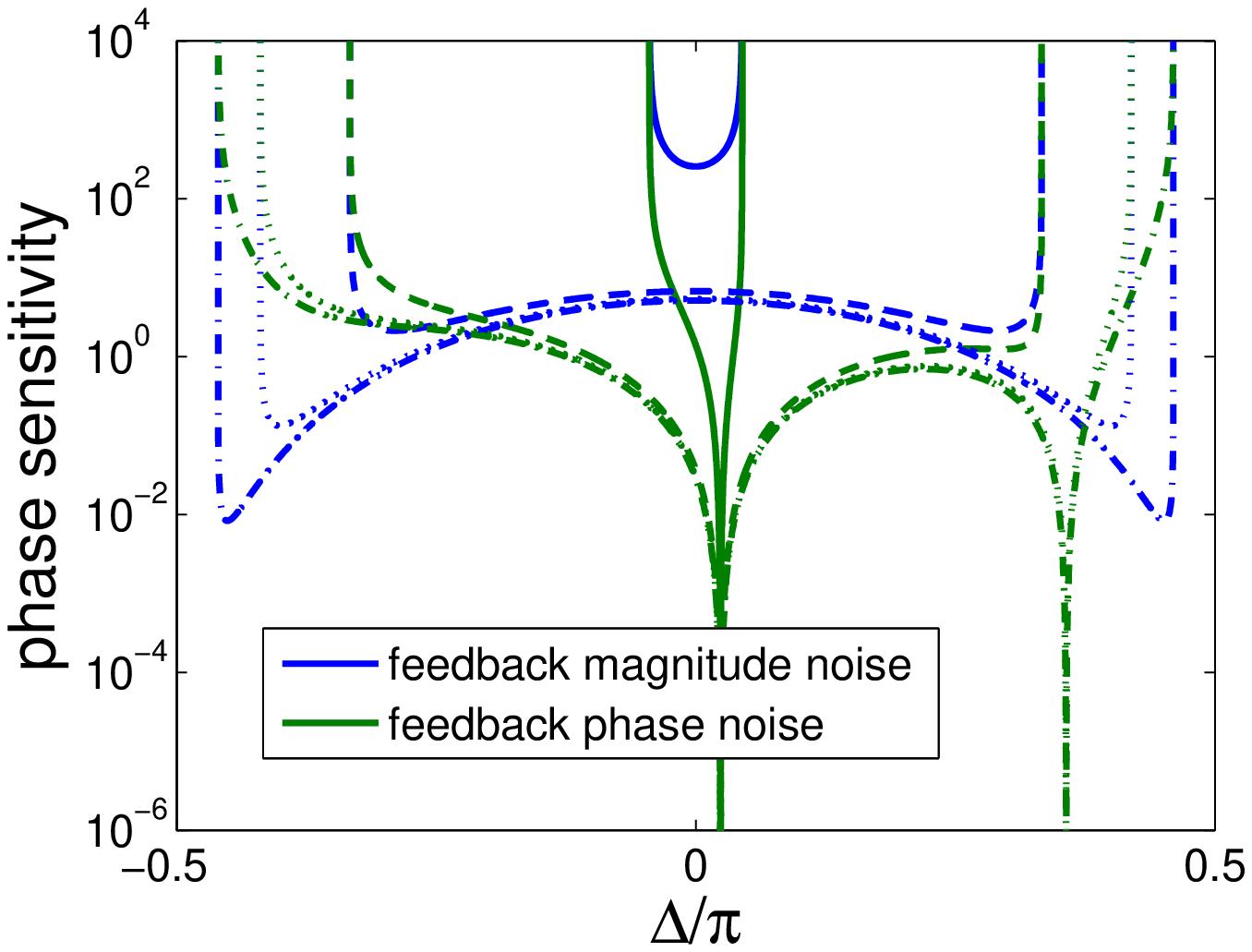}}
   \end{center}
   \caption{\label{fig3} (Color online) (a) Response of an oscillator driven by a non-saturated amplifier for $s=3$ and $k=2$. (b) The phase noise sensitivity coefficients $P_{R}^{2}$ (blue) and $P_{I}^{2}$ (green) for the same amplifier gain values. Note the cancellation of amplifier phase noise at small amplitudes for $G=1.01$.}
\end{figure}

Away from the saturated limit, some level of phase diffusion persists, but we can reduce it by parameter tuning. The long time drift of the phase due to the noise terms is given by solving \cite{Kenig_PRE}
\begin{equation}\label{}
    \dot{\phi}=P_R\,\Xi_{R}+ P_I\,\Xi_{I},\label{stochastic phase}
\end{equation}
where $P_{R},P_{I}$ determine the effect on the phase of each noise term appearing in Eq.~(\ref{vector_SDEs}) and are given by the component of the corresponding noise vector along $V_\perp$
\begin{eqnarray}\label{phaseSensitivity}
  P_R &=& V_\perp \cdot \left( \begin{array}{cc} \tfrac{1}{2}\cos \Delta \\ \tfrac{1}{2}a_0^{-1} \sin \Delta \end{array} \right), \\
  P_I &=& V_\perp \cdot \left( \begin{array}{cc} -\tfrac{1}{2} \sin \Delta \\ \tfrac{1}{2}a_{0}^{-1}\cos \Delta \end{array} \right). \nonumber
\end{eqnarray}
The spectrum of the oscillator phase noise depends on the spectral properties of the noise sources $\Xi_{R},\Xi_{I}$, but the dependence of the overall intensity of the noise on the oscillator properties is determined by the quantities $P_{R}, P_{I}$.
Figure \ref{fig3} (b) shows the situation for some typical choices of the amplifier gain parameter $G$, with $s=3$ and $k=2$.  Plotted are the {\it phase noise sensitivity coefficient} $P_R^2$ and $P_I^2$ {\it vs.}\ the phase shift parameter $\Delta$.  We see that although the coefficient $P_R^2$ remains positive, $P_I^2$ can be set to zero by tuning $\Delta$.  This means that, even in the unsaturated case, it is possible to eliminate the ``direct" contribution to phase diffusion.

The noise elimination condition $P_I=0$ can be written in a useful and compact way, namely
\begin{equation}
\frac{d\Omega_0}{d\Delta} = 0 .
\label{Cross_condition}
\end{equation}
This follows from a direct calculation (see Eq.~(\ref{Vperp}))
\begin{eqnarray}
\frac{d\Omega_0}{d\Delta} &=&   \frac{\partial \Omega_0}{\partial a}  \frac{da}{d\Delta} +  \frac{ \partial \Omega_0}{\partial \Delta} = - \frac{\left(\frac{\partial f_\Phi(a_0)}{\partial a}\right)}{\left(\frac{\partial f_a(a_0)}{\partial a}\right)}  \frac{\partial f_a(a_0)}{\partial \Delta} + \frac{\partial f_\Phi(a_0)}{\partial \Delta} \nonumber\\
&=&\frac{g(a_0)}{2}\left[\frac{\left(\frac{\partial f_\Phi}{\partial a}\right)}{\left(\frac{\partial f_a}{\partial a}\right)}\sin \Delta + \frac{\cos \Delta}{a_0}\right]=g(a_0)P_I.
\label{PI_Cross_condition}
\end{eqnarray}
Evaluating $d\Omega_{0}/{d\Delta}$ for the explicit form of $g(a_0)$, and the oscillation amplitude (\ref{a0_solution})  yields
\begin{eqnarray}
P_I &=&\frac{1}{g(a_0)}\bigg(-\frac{3}{4G^2}s^2 \tan \Delta  (G \cos \Delta )^k \left((G \cos \Delta
   )^k-1\right)^{\frac{2}{k}-1}\nonumber\\
   &+&\frac{1}{2\cos^2\Delta} \bigg).
\label{P_I general}
\end{eqnarray}
Interestingly, for $k=2$ we get the \emph{same} condition for noise elimination as in the saturated regime. The condition is {\it independent of the gain $G$}, and so can be satisfied even when the gain is chosen so that the drive level on the resonator is well below that needed to drive it beyond the Duffing critical amplitude, as we now show.

\subsection*{Small Amplitude Limit}

Although Eq.~(\ref{P_I general}) gives the general result for $P_{I}$ for all amplitudes of oscillations, it is instructive to investigate the small amplitude limit more explicitly by evaluating $d\Omega_{0}/d\Delta$ for small amplitudes.  From Eq.~(\ref{noisefreearray}), the periodic orbit amplitude $a_0$ and frequency $\Omega_{0}$ satisfy
\begin{equation}
a_0 = g(a_0) \cos \Delta,
\label{a}
\end{equation}
and
\begin{equation}
\Omega_{0} = \frac{3}{8} a_0^2 + \frac{1}{2} \tan \Delta,
\label{Omega}
\end{equation}
so that
\begin{equation}
\frac{d\Omega_{0}}{d\Delta}=\frac{3}{8}\frac{da_0^{2}}{d\Delta}+\frac{1}{2}\sec^{2}\Delta.
\label{domega dDelta}
\end{equation}
For small input amplitudes, we  expect the amplifier gain function to be given by a Taylor expansion
\begin{equation}
g(a_0)=Ga_0-\beta a_0^{3}+\cdots \quad ,
\label{Taylor}
\end{equation}
where only odd powers are present since $a_{0}\to -a_{0}$ corresponds simply to a $\pi$ phase shift of the periodic input signal. The coefficient $\beta$ gives the leading order nonlinearity of the amplifier gain \footnote{Note that the Rapp Model Eq.~(\ref{Rapp}) is consistent with the Taylor expansion with $\beta\ne 0$ only for $k=2$.}. The oscillator amplitude is small and then the amplitude is
\begin{equation}\label{smallAmp}
a_0^{2}\simeq\frac{1}{\beta}\left(G-\sec\Delta\right),
\end{equation}
so that
\begin{equation}
\frac{da_0^{2}}{d\Delta}=-\frac{1}{\beta}\sec\Delta\tan\Delta.
\end{equation}
This gives
\begin{equation}
\frac{d\Omega_{0}}{d\Delta}=-\frac{3}{8\beta}\sec\Delta\tan\Delta+\frac{1}{2}\sec^{2}\Delta.\label{cubic noise}
\end{equation}
There are values giving $d\Omega_{0}/d\Delta=0$ and so zero phase noise sensitivity $P_I$ at
\begin{equation}
\sin \Delta=\frac{4}{3}\beta ,
\end{equation}
which has solutions if $\beta<3/4$. Translating to physical quantities this means that the nonlinearity of the amplifier gain $g(a_0)$ must be such that the resonator amplitudes sufficient to change the gain by a significant amount are comparable to the Duffing bifurcation amplitude ($a_0\sim 1$). However the resonator does not have to be driven to this amplitude in the closed-loop oscillator to achieve the zero-noise points, since Eq.~(\ref{cubic noise}) depends on the value the gain curve curvature and feedback phase, but not the level of the drive.

We can understand the phase noise elimination at the special operating phases in terms of the cancellation between the two terms on the right hand side of Eq.~(\ref{domega dDelta}). The second term $\tfrac{1}{2}\sec^{2}\Delta$ comes from the frequency dependence of the oscillator on the feedback phase, and is present even for a linear resonator. The conventional route to reducing the phase noise sensitivity is to make the effect of this term small by increasing the $Q$ of the resonator (this would be seen by writing Eq.~(\ref{domega dDelta}) in terms of unscaled variables). Instead, with a nonlinear resonator, this term can be cancelled using the dependence of the resonator frequency on the amplitude of oscillation, and then the feedback phase dependence of this amplitude. Note that $da_0^{2}/d\Delta$ can be of order unity, sufficiently large to cancel the second term, even for small $a_0^{2}$: this is a result of the $a_0\to 0$ limit being the bifurcation point for the onset of the limit cycle oscillations.

\section{Closed loop parameter sweeps}

Two recent papers have presented experimental results showing parameter scans of oscillators based on high-$Q$ resonators driven into their nonlinear regime \cite{Kenny,HRL_experiment}. A major interest of these works is to trace out the characteristic driven resonator ``Duffing'' curve, showing multiple solutions for the amplitude and phase of the driven oscillations for a given driving frequency, over some frequency range and for sufficiently large drive amplitudes. In the closed loop configuration the phase shift $\Delta$ of the feedback is the natural control parameter. In addition to tuning $\Delta$, these experiments simultaneously tune the amplifier characteristics to maintain the drive level on the resonator at a fixed value. Since the drive level is fixed, the amplitude-frequency variation as $\Delta$ is tuned follows the resonator response curve for that drive level. An advantage of this method of measuring the resonator response curve is that branches of the curve corresponding to unstable solutions in the open loop configuration are stabilized by the closed-loop feedback, so the whole of the response curve can be measured. Mimicking this protocol allows us to present our results on the phase noise in a particularly graphic way.
A first important statement is the perhaps obvious one that the noise properties at particular parameter values cannot depend on the nature of the parameter sweep that led to those values. In particular, although the experimentalist can measure $\Omega_{0}(\Delta)$ (how the oscillator frequency depends on the phase shift) in these constant drive sweeps, the derivative of this curve $d\Omega_{0}/d\Delta$ does \emph{not} give the correct function to evaluate the phase noise sensitivity coefficient $P_{I}$ in (\ref{PI_Cross_condition}), since the derivative in this equation must be taken at constant system parameters, and in the constant drive sweep the system parameters characterizing the amplifier are changed. This is unlike the case of the saturated amplifier, where the drive level on the resonator is indeed fixed by the saturation.

\begin{figure}[]
\begin{center}
      \subfigure[]{
   \includegraphics[width=0.31\columnwidth]{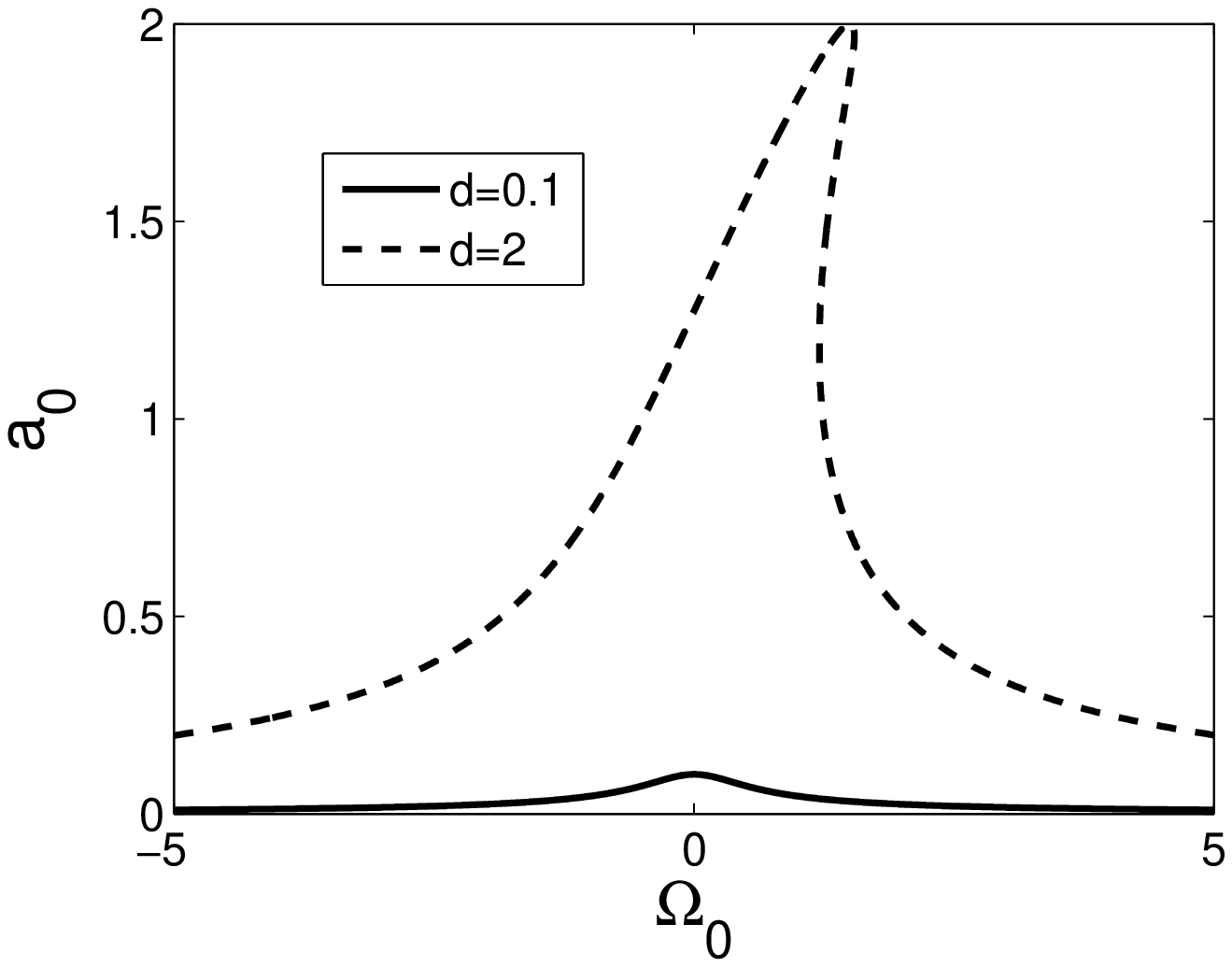}}
  \subfigure[]{
   \includegraphics[width=0.31\columnwidth]{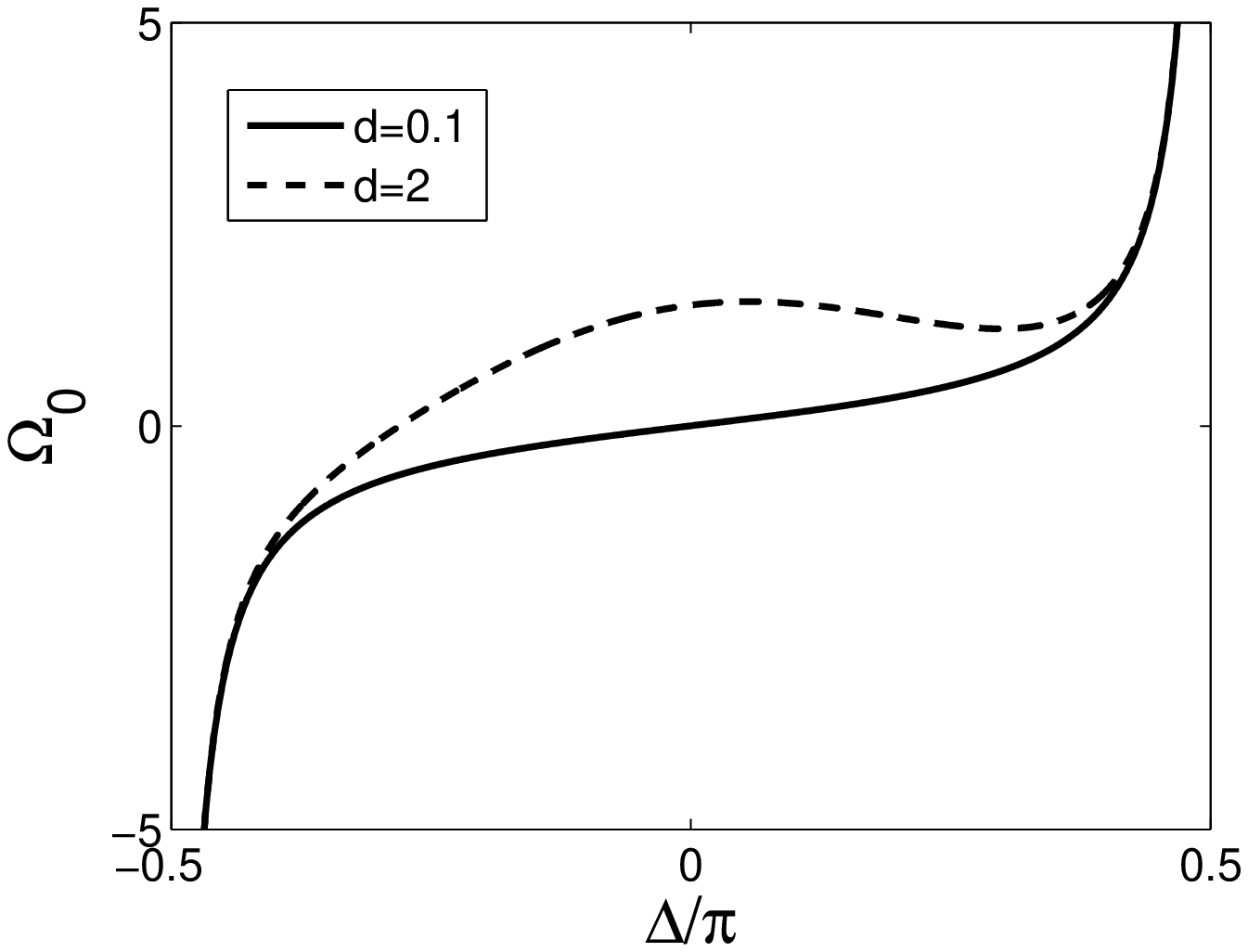}}
     \subfigure[]{
   \includegraphics[width=0.31\columnwidth]{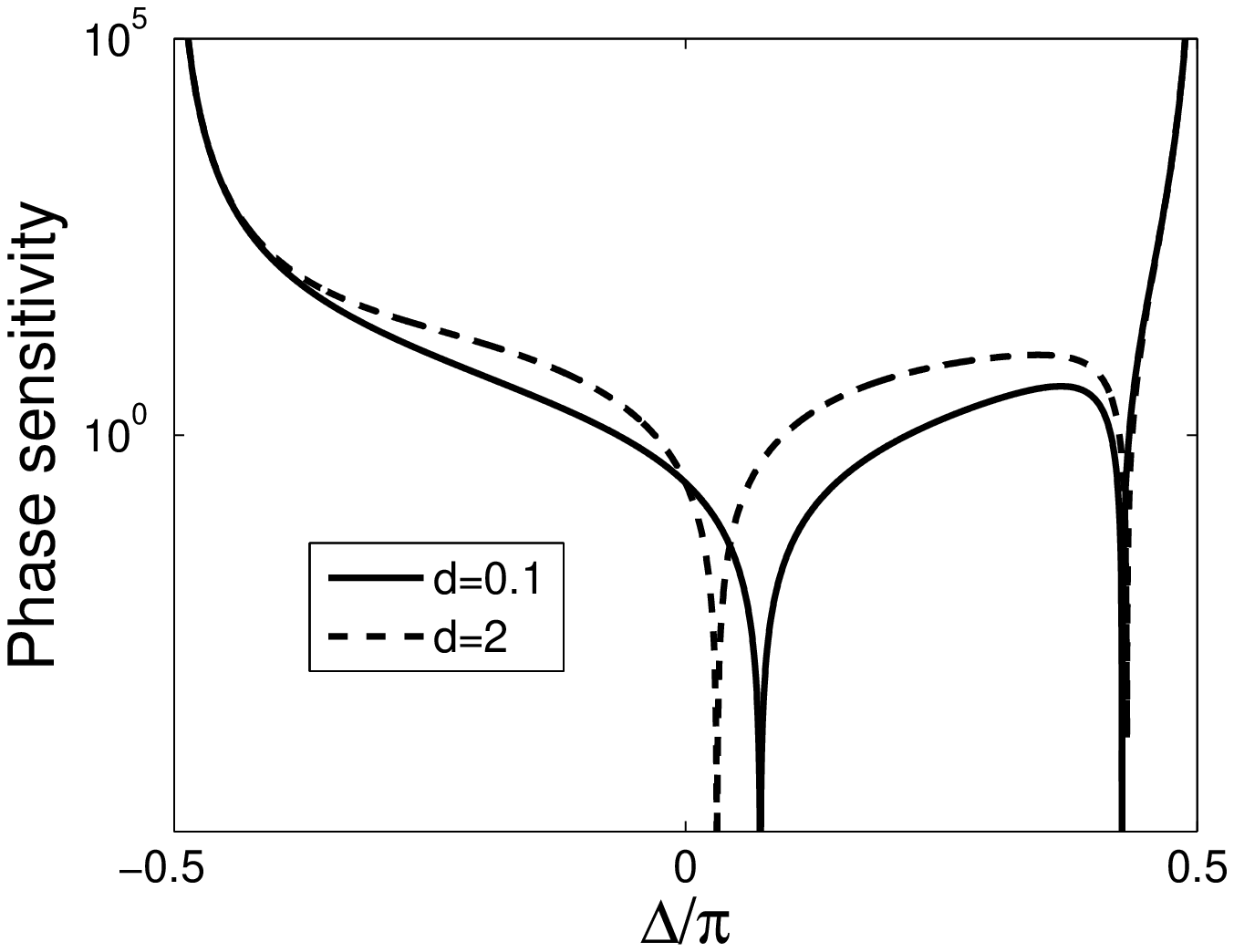}}
   \end{center}
   \caption{\label{fig4} Oscillation characteristics for constant feedback sweeps for $k=2$, $a_s\simeq1.73$, and two feedback drive levels $d=0.1$ (dashed) and $d=2$ (solid): (a) oscillation amplitude {\it vs.}~oscillation frequency; (b) oscillation frequency {\it vs.}~phase shift; (c\,) scaled phase noise sensitivity coefficient $(P_I\cdot d)^2$ {\it vs.}~phase shift.
}
\end{figure}

In Fig.~\ref{fig4} we plot the oscillator characteristics for two fixed drive level $\Delta$ sweeps following a protocol analogous to the one in Ref.~\cite{HRL_experiment}. To implement this it is convenient to write the amplifier gain function as
\begin{equation}\label{Rapp2}
g(a)=\frac{Ga}{\left[1+\left(\frac{a}{a_{s}}\right)^{k}\right]^{1/k}},
\end{equation}
introducing the ``shoulder amplitude'' $a_{s}=s/G$ giving, roughly, the input amplitude at which the nonlinearity of the amplifier becomes strong. Then the protocol of Ref.~\cite{HRL_experiment} corresponds to tuning the gain $G(\Delta)$, whilst holding $a_{s}$ fixed, so that the feedback drive strength $d=g(a)$ remains fixed at a chosen value as the oscillation amplitude $a_{0}$ changes with $\Delta$.  The phase noise sensitivity is calculated from Eq.~(\ref{P_I general}) using $s=Ga_{s}$ and calculating $G(\Delta)$ from the solution for the oscillator amplitude at drive level $d$ given by $a_{0}=d\cos\Delta,d=g(a_{0})$. Note that there are values of $\Delta$ yielding zero phase noise sensitivity $P_{I}=0$ in both cases, even though for the smaller drive level $d=0.1$ the resonator is driven far below the onset of nonlinearity, so that $a_{0}(\Omega_{0})$ follows the linear resonator response curve. Also, these special points are not associated with zeros in the slope of the $\Omega_{0}(\Delta)$ curves, as shown in (b) and (c).

\section{Total Oscillator Phase Noise and Power Spectrum}

\begin{figure}[]
  \begin{center}
  \includegraphics[width=1\columnwidth]{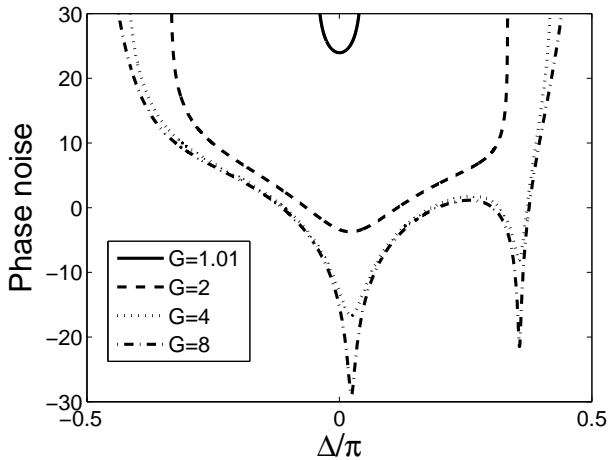}
  \caption{The total sensitivity to both quadratures of amplifier noise, as given by the expression (\ref{aprxPhaseNoise}), for $s=3$ and $k=2$. As the gain level grows the phase noise approaches the saturated amplifier behavior having two zero phase noise points \cite{Kenig_PRE}.}\label{fig5}
  \end{center}
\end{figure}

We have presented results for the phase noise elimination due to the component of amplifier noise in the phase quadrature $\Xi_{I}$, but, as described above, for an unsaturated amplifier there will usually in addition be noise in the magnitude quadrature $\Xi_{R}$ which cannot be eliminated by any choice of $\Delta$ (see Fig.~\ref{fig3}). Thus the ability to reduce the total noise depends on the relative strength of these two components.  Returning to the specific model which led to Eq.~(\ref{XiRandXiI}), the stochastic evolution of the phase Eq.~(\ref{stochastic phase}) becomes
\begin{equation}\label{}
    \dot{\phi}=\bar P_R\bar \xi_{R}+\bar P_I\bar\xi_{I},
\end{equation}
with $\bar P_{I}=g(a_0)P_I/a_0$, and $\bar P_{R}=g'(a_0)P_R$, the phase sensitivities scaled by the relative noise intensities derived from the amplifier-noise model. For $\bar\xi_{R},\bar\xi_{I}$ uncorrelated and equal intensity the total
phase sensitivity to the noise is given by
\begin{equation}
P_{\text{eff}}^{2}={\bar P_{R}^{2}+\bar P_{I}^{2}} ,
\end{equation}
which is plotted in Fig.~\ref{fig5}. We have chosen the value $k=2$ and a saturation level $s=3$, which means that for the largest amplifier output the resonator is driven at a strength about twice the Duffing critical value. The plot shows the effective noise sensitivity coefficient for four gain values $G$. For $G=1.01$ there is only a small range of $\Delta$ giving sustained oscillations and the amplitude of oscillations always remains small (cf.\ Fig.~\ref{fig3}(a)): in this case the noise in the magnitude quadrature of the amplifier output is comparable to the noise in the phase quadrature, so that little effect of the eliminating the phase component is seen. For $G=2$ the oscillation amplitude rises to about 2.5, sufficient to probe the nonlinear region of the amplifier characteristics (see Fig.~\ref{fig1}), and there is some appearance of noise quenching around $\Delta=0$. Already for a gain level $G=4$ there is significant noise quenching, and for $G=8$ the noise curve is close to the value for a saturated amplifier.

To connect these results with the noise spectrum of the oscillator we need to make assumptions on the properties of the noise sources. For stationary, uncorrelated noise sources $\langle\bar \xi_i(T)\bar \xi_j(T')\rangle=R_j(T-T')\delta_{ij}$, the growth of the phase variance $V(\tau)=\langle[\phi(\tau+T)-\phi(T)]^2\rangle$ with time is given by \cite{Kenig_PRE,Demir02}
\begin{eqnarray}\label{var}
    V(\tau)&=&\frac{4\sum_{i}\bar P_{i}^2}{\pi}\int_{0}^{\infty}S_{i}(\Omega)\left[\frac{\sin(\Omega\tau/2)}{\Omega}\right]^2d\Omega,
\end{eqnarray}
with $S_{i}$ the Fourier transform of the noise correlation function $S_{i}(\Omega)={\cal F}[R_{i}(T)]$.
The conventionally quoted phase noise in dBc/Hz at offset frequency $\omega_m$ is then approximated by \cite{Kenig_PRE,Demir02}
\begin{equation}\label{}
    L(\omega_m)=10\log_{10}\left({\cal F}\left[e^{-V(t\omega_0/Q) /2}\right]\right).
\end{equation}
For weak noise, the exponent can expanded to first order, and the phase noise for a given frequency offset is
\begin{equation}\label{aprxPhaseNoise}
    L(\omega_{m})=10\log_{10}\left(P_{\text{eff}}^{2}\right)+C(\omega_{m}),
\end{equation}
where the additive term $C$ depends on the offset frequency, as well as the oscillator frequency and the noise strengths.
Thus Fig.~\ref{fig5} directly shows how the oscillator noise at some chosen offset frequency depends on the feedback phase and amplifier parameters, up to an overall additive constant (on the log scale).

\section{Summary and Conclusions}
In this paper we have investigated the conditions for complete elimination of phase noise in self-sustained oscillators due to fluctuations in the feedback drive. We have shown that the possibility for phase noise elimination lies in the ideal limit of a saturated amplifier, where fluctuations in the feedback magnitude are quenched. In this limit the oscillator response curve reproduces the open loop resonator response curve, and the remaining fluctuations in the
phase of the feedback can be eliminated by tuning the oscillator to the turning points of the resonator Duffing curve. Away from this limit, amplifier noise consists of fluctuations in both the magnitude and the phase of the feedback, a situation which can be represented in phase space by surrounding the operational point with a random noise ellipse that can not be eliminated. By considering an amplitude dependant feedback function with a noisy input, we obtain the total phase noise of the oscillator, and recover complete noise elimination at the large amplitude limit.

We show that the possibility for elimination of fluctuations in the feedback phase does not rely on the large amplitude restriction, and in the small amplitude limit this ability depends on the nonlinearity of the amplifier rather than the oscillation amplitude at the operational point. We demonstrate complete elimination of fluctuations in the feedback phase for a feedback level much lower than the critical level for nonlinear Duffing response. This effect has a large impact in situations where fluctuations in the phase of the feedback dominate.
\acknowledgments
The authors wish to thank Stephen Wandzura and Harris (Chip) Moyer for helpful discussions.
This research was supported by DARPA through the DEFYS program.

\bibliography{phaseNoiseOfUnsaturatedOscillators}

\end{document}